%% file: anomaly_med-2014.tex
\documentclass[letterpaper, 10 pt, conference]{ieeeconf}

\IEEEoverridecommandlockouts

\makeatletter
\def\ps@headings{%
\def\@oddhead{\mbox{}\scriptsize\rightmark \hfil \thepage}%
\def\@evenhead{\scriptsize\thepage \hfil \leftmark\mbox{}}%
\def\@oddfoot{}%
\def\@evenfoot{}}
\makeatother

\usepackage{amsmath}
\usepackage{amstext}
\usepackage{amssymb}
\usepackage{amsfonts}
\usepackage{pstricks}
\usepackage{theorem}
\usepackage[hang]{subfigure}
\usepackage{psfrag}
\usepackage{float}
\usepackage{floatflt}
\usepackage{mathrsfs}
\usepackage{multirow}
\usepackage{epsfig}
\usepackage{verbatim}

\input{mymacros_2cln}

\input{lpsymbols}

\floatstyle{boxed}
\newfloat{algorithm}{htp}{lop}
\floatname{algorithm}{Algorithm}

\def\1{{\mathbf 1}}

\def\msS{{\mathcal S}}
\def\s{{\mathbf s}}
\def\x{{\mathbf x}}

\def\f{{\mathbf f}}
\def\msF{{\mathcal F}}
\def\msX{{\mathcal X}}

\def\bsigma{{\boldsymbol \sigma}}
\def\brho{{\boldsymbol \rho}}

\def\bmu{{\boldsymbol \mu}}

\def\bnu{{\boldsymbol \nu}}

\def\bPi{{\boldsymbol \Pi}}

\def\bxi{{\boldsymbol \xi}}
\def\balpha{{\boldsymbol \alpha}}

\def\s{{\mathbf s}}

\usepackage{algpseudocode} 

\begin{document}

\title{\LARGE \bf Robust Anomaly Detection in Dynamic
  Networks~\authorrefmark{1} \thanks{* Research partially supported by the
    NSF under grants CNS-1239021, IIS-1237022, by the DOE under grant
    DE-FG52-06NA27490, by the ARO under grants W911NF-11-1-0227 and
    W911NF-12-1-0390, by the ONR under grant N00014-10-1-0952, and by the
    NIH/NIGMS under grant GM093147.}  }

\author{Jing Wang\authorrefmark{2}
  \thanks{
    $\dagger$ Division of Systems Engineering, Boston University,
    8 St. Mary's St., Boston, MA 02215, 
    {\tt wangjing@bu.edu.}} and 
Ioannis Ch. Paschalidis\authorrefmark{3}
  \thanks{
    $\ddagger$ Department of Electrical and Computer Engineering and
  Division of Systems Engineering, Boston University,
    8 St. Mary's St., Boston, MA 02215,
    {\tt yannisp@bu.edu}, {\tt http://ionia.bu.edu/.}}
}

\maketitle
\global\long\def\Pr{\ensuremath{\mbox{Pr}}}
\global\long\def\bPth{\bP_{\bth}}
\global\long\def\testSize#1{\alpha^{#1}\left(\boldsymbol{\theta}\right)}
\global\long\def\testMPower#1{\beta^{#1}\left(\boldsymbol{\theta}\right)}
\global\long\def\msS{{\mathcal{S}}}
\input{wj_macros.tex}

\begin{abstract}
  We propose two robust methods for anomaly detection in dynamic networks
  in which the properties of normal traffic are time-varying.  We formulate
  the robust anomaly detection problem as a \emph{binary composite
  hypothesis testing problem} and propose two methods: a \emph{model-free}
  and a \emph{model-based} one, leveraging techniques from the theory of
  large deviations. Both methods require a family of Probability Laws (PLs)
  that represent normal properties of traffic.  We devise a two-step
  procedure to estimate this family of PLs. We compare the performance of
  our robust methods and their vanilla counterparts, which assume that
  normal traffic is stationary, on a network with a diurnal normal pattern
  and a common anomaly related to data exfiltration.  Simulation results
  show that our robust methods perform better than their vanilla
  counterparts in dynamic networks.
\end{abstract}

\vspace{-0.36cm}

\section{Introduction}

A network anomaly is any potentially malicious traffic sequence that has
implications for the security of the network. Although automated online
traffic anomaly detection has received a lot of attention, this field is far
from mature.

Network anomaly detection belongs to a broader field of system anomaly
detection whose approaches can be roughly grouped into two classes:
\emph{signature-based anomaly detection}, where known patterns of past
anomalies are used to identify ongoing
anomalies~\cite{roesch1999snort,paxson1999bro}, and \emph{change-based
anomaly detection} that identifies patterns that substantially deviate from
normal patterns of operations~\cite{barford2002signal,Lu2009,pas-sma-ton-09}.
\cite{lippmann2000evaluating} showed that the detection rates of systems
based on pattern matching are below 70\%. Furthermore, such systems cannot
detect \emph{zero-day attacks}, i.e., attacks not previously seen, and need
constant (and expensive) updating to keep up with new attack signatures. In
contrast,\emph{ change-based anomaly detection} methods are considered to be
more economic and promising since they can identify novel attacks. In this
work we focus on \emph{change-based anomaly detection} methods, in particular
on \emph{statistical anomaly detection} that leverages statistical methods.

Standard \emph{statistical anomaly detection} consists of two steps. The
first step is to learn the ``normal behavior'' by analyzing past system
behavior; usually a segment of records corresponding to normal system
activity. The second step is to identify time instances where system
behavior does not appear to be normal by monitoring the system
continuously.

For anomaly detection in networks, \cite{pas-sma-ton-09} presents
two methods to characterize normal behavior and to assess deviations
from it based on the \emph{Large Deviations Theory} (LDT)~\cite{deze2}.
Both methods consider the traffic, which is a sequence of flows, as a
sample path of an underlying stochastic process and compare current
network traffic to some reference network traffic using LDT. One method,
which is referred to as the \emph{model-free} method, employs the method of
types~\cite{deze2} to characterize the type (i.e., empirical measure) of
an independent and identically distributed~(i.i.d.) sequence of network
flows. The other method, which is referred to as the \emph{model-based}
method, models traffic as a \emph{Markov Modulated Process.} Both
methods rely on \emph{a stationarity assumption} postulating that the
properties of normal traffic in networks do not change over time.

However, the \emph{stationarity assumption} is rarely satisfied in
contemporary networks~\cite{Neal-2010}. For example, Internet traffic is
subject to weekly and diurnal variations~\cite{thompson1997wide,King2013}.
Internet traffic is also influenced by macroscopic factors such as important
holidays and events~\cite{Sandvine2013}. Similar phenomena arise in local area
networks as well.  We will call a network \emph{dynamic} if its traffic
exhibits time-varying behavior.

The challenges for anomaly detection of dynamic networks are two-fold.  First,
the methods used for learning the ``normal behavior'' are usually quite
sensitive to the presence of non-stationarity.  Second, the modeling and
prediction of multi-dimensional and time-dependent behavior is hard.

To address these challenges, we generalize the vanilla \emph{model-free} and
\emph{model-based} methods from \cite{pas-sma-ton-09} and develop what we call
the \emph{robust model-free} and the \emph{robust model-based} methods. The
novelties of our new methods are as follows. First, our methods are robust and
optimal in the generalized Neyman-Pearson sense.  Second, we propose a
two-stage method to estimate Probability Laws (PLs) that characterize normal
system behaviors. Our two-stage method transforms a hard problem (i.e.,
estimating PLs for \emph{multi-dimensional} data) into two well-studied
problems: $(i)$ estimating \emph{one-dimensional} data parameters and $(ii)$
the \emph{set cover} problem. Being concise and interpretable, our estimated
PLs are helpful not only in anomaly detection but also in understanding normal
system behavior.  
% Moreover, our two-stage method is suitable for distributed
% computation.

The structure of the paper is as
follows. Sec.~\ref{sec:Binary-Composite-Hypothesis} formulates system anomaly
detection as a binary composite hypothesis testing problem and proposes two
robust methods. Sec.~\ref{sec:Network-Anomaly-Detection} applies the methods
presented in
Sec.~\ref{sec:Binary-Composite-Hypothesis}. Sec.~\ref{sec:Network-Simulation}
explains the simulation setup and presents results from our robust methods as
well as their vanilla counterparts.  Finally, Sec.~\ref{sec:Conclusions}
provides concluding remarks.

\section{Binary composite hypothesis
testing\label{sec:Binary-Composite-Hypothesis}} 

We model the network environment as a stochastic process and estimate its
parameters through some reference traffic (viewed as sample paths). Then
the problem of network anomaly detection is equivalent to testing whether a
sequence of observations $\calG=\lt\{g^{1},\ldots,g^{n}\rt\}$ is a sample
path of a discrete-time stochastic process
$\scrG=\lt\{G^{1},\ldots,G^{n}\rt\}$ (hypothesis $\calH_0$). All random
variables $G^{i}$ are discrete and their sample space is a finite alphabet
$\Sigma=\lt\{\sigma_{1},\sigma_{2},\dots,\sigma_{|\Sigma|}\rt\}$, where
$|\Sigma|$ denotes the cardinality of $\Sigma$. All observed symbols
$g^{i}$ belong to $\Sigma$, too.  This problem is a \emph{binary composite
hypothesis testing problem}.  Because the joint distribution of all random
variables $G^i$ in $\scrG$ becomes complex when $n$ is large, we propose
two types of simplification.  

\subsection{A model-free method\label{sub:A-Model-Free-Approach}}

We propose a \emph{model-free} method that assumes the random variables
$G^{i}$ are i.i.d. Each $G^i$ takes the value $\sigma_{j}$ with
probability $p_{\theta}^{F}(G^i=\sigma_{j})$, $j=1,\ldots,|\Sigma|$, which
is parameterized by $\theta\in\Omega$. We refer to the vector 
$\bp_{\theta}^{F}=(p_{\theta}^{F}(G^i=\sigma_{1}),\ldots,
p_{\theta}^{F}(G^i=\sigma_{|\Sigma|}))$ as the {\em model-free}
Probability Law (PL) associated with $\theta$. Then the family of
\emph{model-free} PLs $\scrP^{F}=\left\{
\bp_{\theta}^{F}:\theta\in\Omega\right\} $ characterizes the stochastic
process $\scrG$.

To characterize the observation $\calG$, let
\begin{equation}
\scrE_{F}^{\calG}\lt(\sigma_{j}\rt)=\frac{1}{n}\sum_{i=1}^{n}
\mathbf{1}\lt(g^{i}=\sigma_{j}\rt),  
\qquad j=1,\ldots,|\Sigma|, \label{eq:mf-em}
\end{equation}
where $\1(\cdot)$ is an indicator function. Then, an estimate for
the underlying \emph{model-free} PL based on the observation $\calG$ is
$\boldsymbol{\calE}_{F}^{\calG}=\left\{
\scrE_{F}^{\calG}(\sigma_{j}):\ j=1,\dots,|\Sigma|\right\} $, which is
called the \emph{model-free} empirical measure of $\calG$.

Suppose $\bmu=(\mu(\sigma_{1}), \ldots, \mu(\sigma_{|\Sigma|}))$ is a
\emph{model-free} PL and $\bnu=(
\nu(\sigma_{1}),\ldots,\nu(\sigma_{|\Sigma|}))$ is a \emph{model-free}
  empirical measure. To quantify the difference between $\bmu$ and
  $\bnu$, we define the \emph{model-free} \emph{divergence} between
  $\bmu$ and $\bnu$ as
\begin{equation}
D_{F}(\bnu\Vert\bmu)\triangleq
\sum_{j=1}^{|\Sigma|}\hat{\nu}(\sigma_{j})\log\frac{\hat{\nu}(\sigma_{j})}{\hat{\mu}(\sigma_{j})}, \label{eq:model-free-cross-entropy}
\end{equation}
where $\hat{\nu}(\sigma_{j})=\max(\nu(\sigma_{j}),\varepsilon)$ and
$\hat{\mu}(\sigma_{j})=\max(\nu(\sigma_{j}),\varepsilon),\forall j$ and
$\varepsilon$ is a small positive constant introduced
to avoid underflow and division by zero.

\begin{defi}
\label{def:mf-GHT}(Model-Free Generalized Hoeffding Test). The
\emph{model-free generalized Hoeffding test}~\cite{hoef65} is to reject
$\calH_{0}$ if $\calG$ is in
\[
S_{F}^{*}=\lt\{\calG\mid\inf_{\theta\in\Omega}\,
D_{F}\lt(\bcalE_{F}^{\calG}\Vert\bp_{\theta}^{F}\rt)\geq\lambda\rt\}, 
\]
where $\lambda$ is a detection threshold and $\inf_{\theta\in\Omega}\,
D_{F}\lt(\bcalE_{F}^{\calG}\Vert\bp_{\theta}^{F}\rt)$ 
is referred to as the \emph{generalized model-free divergence} between
$\bcalE_{F}^{\calG}$ and $\scrP^{F}=\left\{
\bp_{\theta}^{F}:\theta\in\Omega\right\}$. 
\end{defi}
A similar definition has been proposed for robust localization in sensor
networks~\cite{mainlocalization}. One can show that this generalized
Hoeffding test is asymptotically (as $n\ra \infty$) optimal in a
generalized Neyman-Pearson sense; we omit the technical details in the
interest of space.

\begin{comment}
This paper applies this definition to the anomaly detection of dynamic
networks, in which $\bcalE_{F}^{\calG}$ characterizes the traffic
$\calG$ that needs to be detected and the family of PLs $\scrP^{F}=\left\{ \bp_{\theta}^{F}:\theta\in\Omega\right\} $
is estimated from the reference traffic $\calG_{ref}$.
\end{comment}
\begin{comment}
For a long trace $\calG$, its empirical measure is {}``close to''
to $\bnu$ with probability that behaves as 
\[
\bP\left[\boldsymbol{\calE}_{F}^{\calG}\approx\bnu\right]\asymp e^{-nD_{F}\lt(\bnu\parallel\bp_{\theta}^{F}\rt)}.
\]
We will refer to exponents $D_{F}\lt(\bnu\parallel\bp_{\theta}^{F}\rt)$
as the \emph{exponential decay rate} of $\bP\left[\boldsymbol{\calE}_{F}^{\calG}\approx\bnu\right]$.
\begin{thm}
\label{thm:model-free-GNP}The model-free generalized Hoeffding test
satisfies the GNP criterion. \end{thm}
\end{comment}

\subsection{A model-based method\label{sub:A-Model-Based-Approach}}

We now turn to the \emph{model-based} method where the random process
$\scrG=\{G^{1},\ldots,G^{n}\}$ is assumed to be a Markov chain. Under
this assumption, the joint distribution of $\scrG$ becomes
$p_{\theta}\left(\scrG=\calG\right) =
p_{\theta}^{B}\left(g^{1}\right)\prod_{i=1}^{n-1}p_{\theta}^{B}\left(g^{i+1}\mid
g^{i}\right)$, where $p_{\theta}^{B}(\cdot)$ is the initial distribution
and $p_{\theta}^{B}\left(\cdot\mid\cdot\right)$ is the transition
probability; all parametrized by $\theta\in\Omega$.

Let $p_{\theta}^{B}\left(\sigma_{i},\sigma_{j}\right)$ be the probability
of seeing two consecutive states $(\sigma_{i},\sigma_{j})$.  We refer to
the matrix $\bP_{\theta}^{B}=\{
p_{\theta}^{B}(\sigma_{i},\sigma_{j})\}_{i,j=1}^{|\Sigma|} $
as the \emph{model-based} PL associated with $\theta\in\Omega$. Then, the
family of \emph{model-based} PLs $\scrP^{B}=\left\{
\bP_{\theta}^{B}:\theta\in\Omega\right\} $ characterizes the stochastic
process $\scrG$.

To characterize the observation $\calG$, let 
\begin{equation}
\scrE_{B}^{\calG}\lt(\sigma_{i},\sigma_{j}\rt)=\frac{1}{n}\sum_{l=2}^{n}
\mathbf{1}\lt(g^{l-1}=\sigma_{i},
g^{l}=\sigma_{j}\rt), i,j=1,\ldots,\lt|\Sigma\rt|. \label{eq:mb-em}   
\end{equation}
We define the \emph{model-based} empirical measure of $\calG$ as the
matrix $\boldsymbol{\calE}_{B}^{\calG}=\{
\scrE_{B}^{\calG}(\sigma_{i},\sigma_{j})\}_{i,j=1}^{|\Sigma|}$.  The
transition probability from $\sigma_{i}$ to $\sigma_{j}$ is simply
$\scrE_{B}^{\calG}\lt(\sigma_{j}|\sigma_{i}\rt)=\frac{\scrE_{B}^{\calG}\lt(\sigma_{i},\sigma_{j}\rt)}{\sum_{j=1}^{\lt|\Sigma\rt|}\scrE_{B}^{\calG}\lt(\sigma_{i},\sigma_{j}\rt)}$. 
 
Suppose
$\boldsymbol{\Pi}=\{\pi(\sigma_{i},\sigma_{j})\}_{i,j=1}^{|\Sigma|}$
is a \emph{model-based} PL and $\mathbf{Q}=\{
q(\sigma_{i},\sigma_{j})\}_{i,j=1}^{|\Sigma|}$ is a
\emph{model-based} empirical measure. Let
$\hat{\pi}(\sigma_{j}|\sigma_{i})$ and $\hat{q}(\sigma_{j}|\sigma_{i})$
be the corresponding transition probabilities from $\sigma_{i}$ to
$\sigma_{j}$. Then, the \emph{model-based divergence} between $\bPi$ and
$\bQ$ is

\begin{equation}
D_{B}\lt(\mathbf{Q}\parallel\boldsymbol{\Pi}\rt)=\sum_{i=1}^{\lt|\Sigma\rt|}\sum_{j=1}^{\lt|\Sigma\rt|}\hat{q}(\sigma_{i},\sigma_{j})\log\frac{\hat{q}(\sigma_{j}|\sigma_{i})}{\hat{\pi}(\sigma_{j}|\sigma_{i})},\label{eq:model-based-cross-entropy}
\end{equation}
where
$\hat{q}(\sigma_{i},\sigma_{j})=\max(q(\sigma_{i},\sigma_{j}),\varepsilon)$, 
$\hat{\pi}(\sigma_{i},\sigma_{j})=\max(\pi(\sigma_{i},\sigma_{j}),\varepsilon)$
for some small positive constant $\varepsilon$ introduced to avoid
underflow and division by zero. Similar to the \emph{model-free} case,
we present the following definition: 
\begin{defi}
\label{def:mb_GHT}(Model-Based Generalized Hoeffding Test). The
\emph{model-based generalized Hoeffding test} is to reject $\calH_{0}$
when $\calG$ is in
\[
S_{B}^{*}=\lt\{\calG\mid\inf_{\theta\in\Omega}\, D_{B}\lt(\boldsymbol{\calE}_{B}^{\calG}\Vert\bP_{\theta}^{B}\rt)\geq\lambda\rt\},
\]
where $\lambda$ is a detection threshold and $\inf_{\theta\in\Omega}\, D_{B}\lt(\boldsymbol{\calE}_{B}^{\calG}\Vert\bP_{\theta}^{B}\rt)$
is referred to as the \emph{generalized model-based divergence} between
$\bcalE_{F}^{\calG}$ and $\scrP^{B}=\left\{
\bP_{\theta}^{B}:\theta\in\Omega\right\} $.
\end{defi}
In this case as well, asymptotic (generalized) Neyman-Pearson optimality
can be established. 
\begin{comment}
For a long trace $\calG$, its empirical measure is {}``close to''
to $\bQ$ with probability that behaves as 
\[
\bP\left[\boldsymbol{\calE}^{\calG}\approx\bQ\right]\asymp e^{-nD\lt(\bnu\parallel\bP_{\theta}\rt)}.
\]
We will refer to exponents $D\lt(\bnu\parallel\bP_{\theta}\rt)$ as
the \emph{exponential decay rate} of the $\bP\left[\boldsymbol{\calE}^{\calG}\approx\bnu\right]$.
\begin{thm}
\label{thm:model-based-GNP}The model-based generalized Hoeffding
test satisfies the GNP criterion.\end{thm}
\end{comment}

\section{Network anomaly
  detection \label{sec:Network-Anomaly-Detection}} 

\begin{figure}
\begin{center}
\includegraphics[width=0.63\columnwidth,height=4cm]{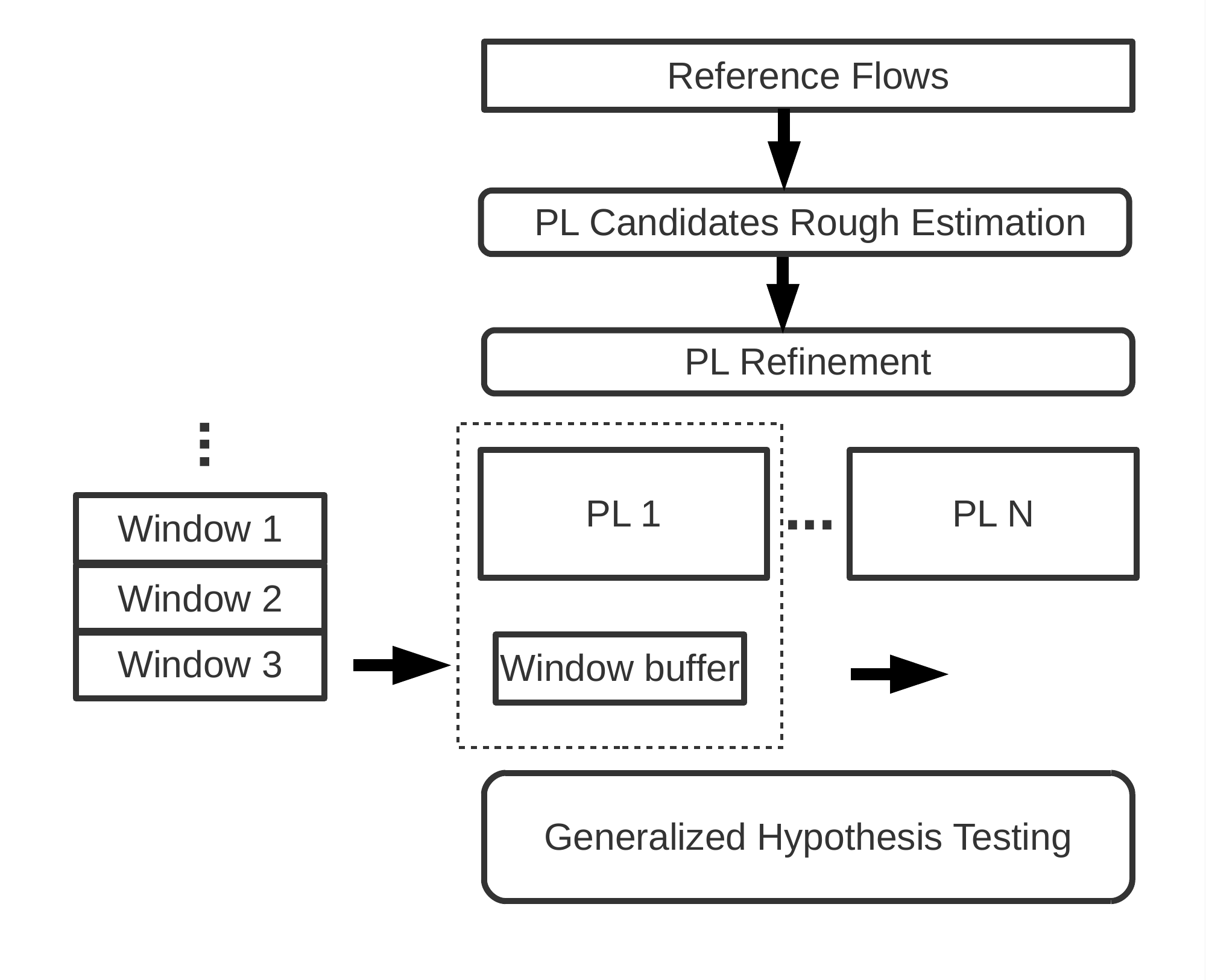}
\vspace{-0.7cm}
\end{center}
\caption{Structure of the algorithms.\label{fig:method-struc}}
\vspace{-0.5cm}
\end{figure}

Fig.~\ref{fig:method-struc} outlines the structure of our robust
anomaly detection methods.  We first propose our feature set
(Sec.~\ref{data-representation}).
We assume that the normal traffic is governed by an underlying stochastic
process $\scrG$.  
% To ease the estimation of the families of PLs for $\scrG$, 
We assume the size of \emph{model-free} and \emph{model-based} PL families
to be finite and propose a two-step procedure to estimate PLs from some
reference data. We first inspect each feature separately to generate a
family of candidate PLs (Sec.\ref{sub:Rough-Estimation-PL}), which is then
reduced to a smaller family of PLs (Sec.~\ref{sub:pl-refinement}). 
For each window, the algorithm applies the \emph{model-free} and
\emph{model-based}
\emph{generalized Hoeffding test} discussed above.

\subsection{Data representation\label{data-representation}}

% Till now, we deliberately formulated problems in a general way.  In fact, the
% only requirement of our methods on data is that each data record should be
% comprised of a time stamp and some additional features. As a result, our
% methods can be used in different applications by choosing appropriate
% features.

In this paper, we focus on \emph{host-based anomaly detection}, a
specific application in which we monitor the incoming and outgoing packets of
a server. We assume that the server provides only one service (e.g., HTTP
server) and other ports are either closed or outside our interests. As a
result, we only monitor traffic on certain port (e.g., port 80 for HTTP
service). For servers with multiple ports in need of monitoring, we can simply
run our methods on each port.

The features we propose for this particular application relate to a flow
representation slightly different from that of commercial vendors
like Cisco NetFlow \cite{netflow}. Hereafter, we will use ``flows'',
``traffic'', and 
``data'' interchangeably.
Let $\calS=\{\bs^{1},\dots,\bs^{|\calS|}\}$ denote the collection
of all packets collected on certain port of the host which is monitored. 
In \emph{host-based anomaly detection}, the server IP is always fixed, thus
ignored.  Denote the user IP address in packet $ $$\s^{i}$ as
$\x^{i}$, whose format will be discussed later. The size of $\s^{i}$ is
$b^{i}\in[0,\infty)$ in bytes and the start time of transmission is
$t^{i}\in[0,\infty)$ in seconds. Using this convention, packet
$\s^{i}$ can be represented
as $(\x^{i},b^{i},t_{s}^{i})$ for all $i=1,\dots,|\calS|$.

% Since packets are many, we consolidate network traffic by grouping
% packets into flows. 
We compile a sequence of packets
% $\s^{1}=(\x^{1},b^{1},t_{s}^{1}),\dots,\s^{m}=(\x^{m},b^{m},t_{s}^{m})$
$\s^{1},\dots,\s^{m}$
with $t_{s}^{1}<\dots<t_{s}^{m}$ into a \emph{flow} $\f=(\x,b,d_{t},t)$
if $\x=\x^{1}=\dots=\x^{m}$ and $t_{s}^{i}-t_{s}^{i-1}<\delta_{F}$ for
$i=2,\dots,m$ and some prescribed $\delta_{F}\in(0,\infty)$.  Here, the
\emph{flow size }$b$ is the sum of the sizes of the packets that
comprise the flow. The \emph{flow duration} is
$d_{t}=t_{s}^{m}-t_{s}^{1}$.  The \emph{flow transmission time} $t$
equals the start time of the first packet of the flow $t_{s}^{1}$. In
this way, we can translate the large collection of packets $\calS$ into
a relatively small collection of flows $\calF$.

Suppose $\msX$ is the set of unique IP addresses in $\msF$. Viewing each IP
as a tuple of integers, we apply typical $K$-means clustering on $\msX$.
For each $\x\in\msX$, we thus obtain a cluster label $k(\x)$. Suppose the
cluster center for cluster $k$ is $\bar{\x}^{k}$; then the distance of $\x$
to the corresponding cluster center is $d_{a}(\x)=d(\x,\bar{\x}^{k(\x)})$,
for some appropriate distance metric.  The cluster label $k(\bx)$ and
distance to cluster center $d_{a}(\bx)$ are used to identify a user IP
address $\bx$, leading to our final representation of a flow as: 
\begin{equation}
\f=(k(\x),d_{a}(\x),b,d_{t},t).\label{eq:flow_distill_def}
\end{equation}
For each $\f$, we quantize $d_{a}(\bx)$, $b$, and $d_{t}$ to discrete
values.  Each tuple of $\left(k(\x),d_{a}(\x),b,d_{t}\right)$ corresponds
to a symbol in $\Sigma=\{1,\dots,K\} \times \Sigma_{d_{a}} \times
\Sigma_{b} \times \Sigma_{d_{t}}$, where $\Sigma_{d_{a}}$, $\Sigma_{b}$ and
$\Sigma_{d_{t}}$ are the quantization alphabets for distance to cluster
center, flow size, and flow duration, respectively.  Denoting by $\bg$ the
corresponding quantized symbol of $\f$ and by $\calG$ the counterpart of
$\calF$, we number the symbols in $\bg$ corresponding to $k(\x)$,
$d_{a}(\x)$, $b$, and $d_{t}$ as features 1, 2, 3, 4.

In our methods, flows in $\calF$ are further aggregated into windows
based on their \emph{flow transmission time}s. A window is a detection
unit that consists of flows in a continuous time range, i.e., the
flows in a same window are evaluated together. Let $h$ be the interval
between the start points of two consecutive time windows and $w_{s}$ be
the window size.

\subsection{Anomaly detection for dynamic networks\label{sub:Candidate-Models}}
For each window $j$, an empirical measure of $\calG_{j}$ is calculated.  We
then leverage the \emph{model-free} and the \emph{model-based} generalized
Hoeffding test (Def.~\ref{def:mf-GHT},\ref{def:mb_GHT}), which require a
set of PLs $\lt\{\bp_{\theta}^{F}:\theta\in\Omega\rt\}$ and
$\lt\{\bPth^{B}:\theta\in\Omega\rt\}$.  We assume $|\Omega|$ to be finite,
and divide our reference traffic $\calG_{ref}$ into segments; the traffic
of each segment is governed by the same PL. The empirical measure of each
segment is then a PL. 

Two flows are likely to be governed by a same PL if they have close
\emph{flow transmission times}. In addition, if the properties of the
normal traffic change periodically, two flows are also likely be governed
by a same PL when the difference of their \emph{flow transmission times}
is close to the period.  Let $t_{p}$ be the period and let $t_{d}$ be
a window size characterizing the speed of change for the normal pattern. We
could divide each period into $\lfloor t_p / t_d \rfloor$ segments with
length $t_d$, and combine corresponding segments of different periods
together, resulting in $\lfloor t_p / t_d \rfloor$ PLs.

In practical networks, the period may vary with time, which makes it hard
to estimate $t_p$ and $t_d$ accurately. To increase the robustness of the
set of estimated PLs to these non-stationarities, we first propose a large
collection of candidates (Sec.~\ref{sub:Rough-Estimation-PL}) and then
refine it (Sec.~\ref{sub:pl-refinement}).

\subsection{Estimation of $t_d$ and $t_p$\label{sub:Rough-Estimation-PL}}
% \begin{figure}
% \centering\includegraphics[width=6.3cm]{figure/QuantizeStateRoughEstimation}
% \vspace{-0.4cm}
% \caption{Histogram of intervals between two consecutive flows with a
%   specific feature quantized to the same discrete
%   value. \label{fig:QuantizeHistogram}}
% \vspace{-0.5cm}
% \end{figure}
% We now present a procedure to generate a set of candidate PLs
% by inspecting each feature separately. 

This section presents a procedure to estimate $t_d$ and $t_p$ by inspecting
each feature separately. Recall that each quantized flow consists of
quantized values of a cluster label, a distance to cluster center, a flow
size and a flow duration, which are called features $1,\dots,4$,
respectively.  We say a quantized flow $\bg$ belongs to \emph{channel}
$a\text{--}b$ if feature $a$ of $\bg$ equals symbol $b$ in quantization
alphabet of feature $a$. We first analyze each channel separately to get a
rough estimate of $t_{d}$ and $t_{p}$. Then, channels corresponding to the
same feature are averaged to generate a combined estimate. 

\begin{figure}
\centering\includegraphics[width=7cm]{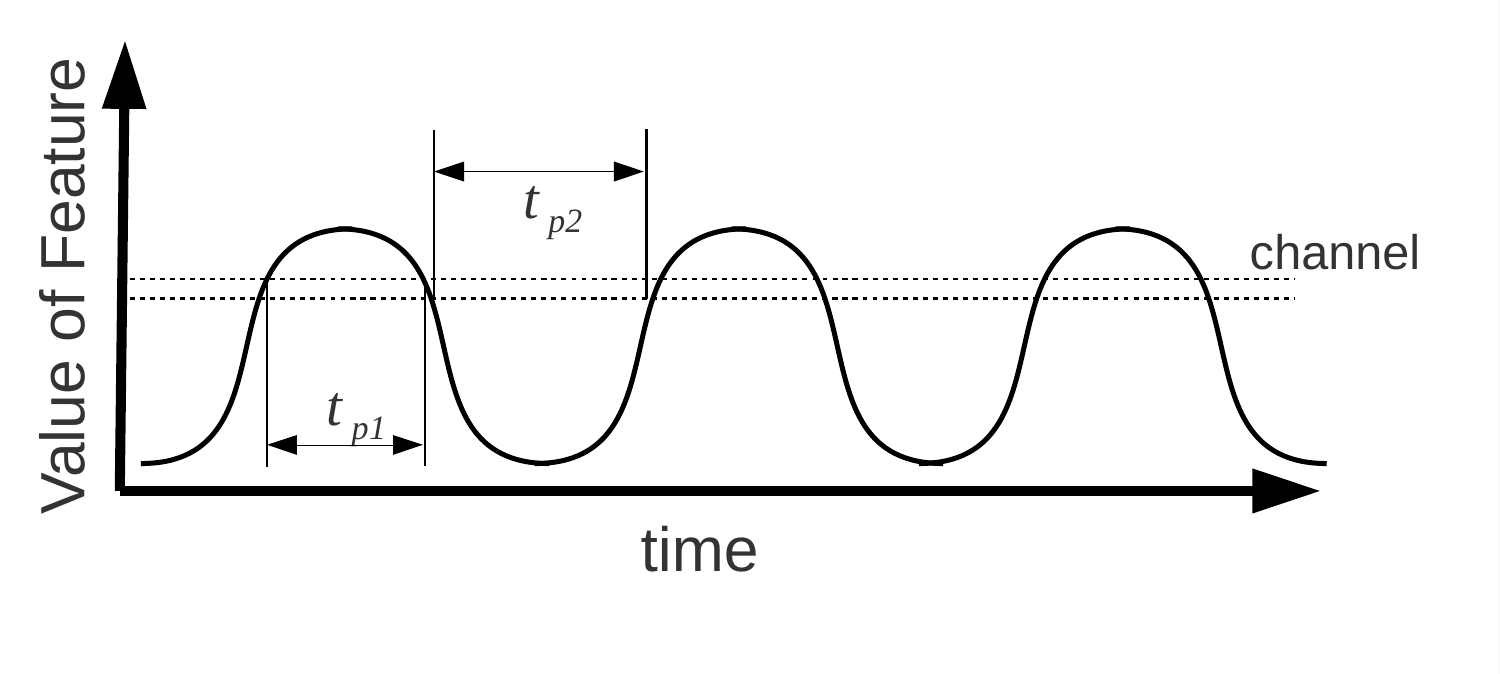}\vspace{-0.6cm} 
\caption{Illustration of the peaks in periodic networks.
% Region 2 of Fig.~\ref{fig:QuantizeHistogram}.
  \label{fig:illustration-peaks-region-3}} 
\vspace{-0.5cm} 
\end{figure}

For all flows in \emph{channel} $a\text{--}b$, we calculate the intervals
between two consecutive flows. Most of the intervals will be very small. If
we divide the interval length to several bins and calculate the histogram,
i.e., the number of observed intervals in each bin. The histogram is
heavily skewed to small interval length.  $t_{d}$ could be chosen to be the
interval length of the first bin (corresponding to the smallest interval
length) whose frequency in the histogram is less than a threshold. In
addition, there may be some large intervals if the feature is periodic.
Fig.~\ref{fig:illustration-peaks-region-3} shows an example of a feature
that exhibits periodicity.  There will be two peaks around $t_{p1}$ and
$t_{p2}$ in the histogram of intervals for flows whose values are between
the two dashed lines.  We can select $t_{p}$ such that
$\left(t_{p1}+t_{p2}\right)/2\thickapprox t_{p}/2$.  There can be a single
or more than two peaks due to noise in the network; in either case, we
choose the average of all peaks as an estimate of $t_{p}/2$.

If no channel of a feature $a$ reports $t_p$, the network is non-periodic
according to the feature $a$.  Otherwise, the estimate of $t_p$ for a
feature $a$ (denoted by $t_p^a$) a is simply the average of all estimates
for channels of the feature $a$. Although the estimate of only one channel
is usually very inaccurate, the averaging procedure helps improve the
accuracy. Similarly, the estimate for $t_d$ for a feature $a$ (denoted by
$t_d^a$) is the average of the estimates for all channels of the feature
$a$.

For each feature $a$, we generate some PLs using the estimate $t_d^a$
and $t_p^a$. In case that some prior knowledge of $t_{d}$ and
$t_{p}$ is available, the family of candidate PLs can include the PLs
calculated based on this prior knowledge.

\subsection{PL refinement with integer programming\label{sub:pl-refinement}}
The larger the family of PLs we use in generalized hypothesis testing, the
more likely we will overfit $\calG_{ref}$, leading to poor results.
Furthermore, a smaller family of PLs reduces the computational cost.  This
section introduces a method to refine the family of candidate PLs.

For simplicity, we only describe the procedure for the \emph{model-free}
method. The procedure for the \emph{model-based} method is similar.
Hereafter, the divergence between a collection of flows and a PL is
equivalent to the divergence between the empirical measure of these flows
and the PL. 

Suppose the family (namely the set) of candidate PLs is the set
$\calP=\{\bp_{1}^{F},\dots,\bp_{N}^{F}\}$ of cardinality $N$. Because no
alarm should be reported for $\calG_{ref}$, or any segment of $\calG_{ref}$,
our \emph{primary objective} is to choose the smallest set
$\scrP^{F}\subseteq\calP$ such that there is no alarm for $\calG_{ref}$. We
aggregate $\calG_{ref}$ into $M$ windows using the techniques of 
% Sec.~\ref{sub:Window-Aggregation} 
Sec.~\ref{data-representation} 
and denote the data in window $i$ as $\calG_{ref}^{i}$. Let
$D_{ij}=D_{F}(\boldsymbol{\calE}^{\calG_{ref}^{i}}\parallel\bp_{j}^{F})$ be
the divergence between flows in window $i$ and PL $j$ 
for $i=1,\ldots,M$ and
$j=1,\ldots,N$. 
We say window $i$ is covered (namely, reported as
normal) by PL $j$ if $D_{ij}\leq\lambda$. With this definition, the
primary objective becomes to select the minimum number of PLs to cover
all the windows.

There may be more than one subsets of $\calP$ having the same cardinality and
covering all windows.  We propose a \emph{secondary objective}
characterizing the variation of a set of PLs. 
% Let
% $N_{j}=\{i:\ D_{ij}\leq\lambda\}$ be the index set of windows covered by
% PL $j$ and denote by $N_{j}^{(1)},\ldots,N_{j}^{(|N_{j}|)}$ the ordered
% elements of $N_{j}$.  
Denote by $\scrD_j$
% $\scrD_{j}=\{ N_{j}^{(i)}-N_{j}^{(i-1)}:\ i=2,\ldots,|N_{j}|\}$ 
the set of intervals between consecutive window covered by PL $j$.
The \emph{coefficient of variation} for PL $j$ is defined as
$c_{v}^{j}=\textsc{Std}(\scrD_{j})/\textsc{Mean}(\scrD_{j})$, where
$\textsc{Std}(\scrD_{j})$ and $\textsc{Mean}(\scrD_{j})$ are the sample
standard deviation and mean of set $\scrD{}_{j}$, respectively.  A smaller
\emph{coefficient of variation} means that the PL is more ``regular.''  
% The
% \emph{secondary objective} is to minimize the sum of \emph{coefficients of
% variation} for selected PLs. 

We formulate PL refinement as a \emph{weighted set cover problem} in which
the weight of PL $j$ is $1+\gamma c_{v}^{j}$, where $\gamma$ is a small
weight for the secondary objective. Let $x_{i}$ be the $0\text{--}1$
variable indicating whether PL $i$ is selected or not; let
$\bx=(x_1,\ldots, x_N)$.  Let $\bA=\{a_{ij}\}$ be an $M\times N$ matrix
whose $(i,j)$th element $a_{ij}$ is set to $1$ if $D_{ij}\leq\lambda$ and
to $0$ otherwise. Here, $\lambda$ is the same threshold we used in
Def.~\ref{def:mf-GHT}. Let $\bc_{v}=(c_{v}^{1},\ldots,c_{v}^{N})$. The
selection of PLs can be formulated as the following integer programming
problem:
\begin{equation}
\begin{array}{rl}
\min & \mathbf{1}^{'}\bx+\gamma\bc_{v}^{'}\bx\\
\text{s.t.} & \bA\bx \geq\mathbf{1},\\
& x_{j} \in \{ 0,1\},\ j=1,\dots,N,
\end{array}\label{eq:lp-PL-selection}
\end{equation}
where $\mathbf{1}$ is a vector of ones. The cost function equals
a weighted sum of the \emph{primary cost} $\mathbf{1}^{'}\bx$ and
the\emph{ secondary cost} $\bc_{v}^{'}\bx$. The first constraint
enforces there is no alarm for $\calG_{ref}^i$ for $\forall i$.
%, $i=1,\ldots,M$. 

\begin{algorithm}
\input{HeuristicRefine.tex}
\caption{Greedy algorithm for PL refinement.\label{alg:solve_PL_refine}}
\end{algorithm}
\begin{comment}
\begin{algorithm}
\input{GreedySolve.tex}
\caption{Greedy Procedure to Solve Set Cover Problem\label{alg:greedy_solve_set_cover}}
\end{algorithm}
\end{comment}
Because (\ref{eq:lp-PL-selection}) is NP-hard, we propose a \emph{heuristic
algorithm} to solve it~(Algorithm~\ref{alg:solve_PL_refine}).
$\textsc{HeuristicRefinePl}$ is the main procedure whose parameters are $\bA$,
$\bc_{v}$, a discount ratio $r<1$, and a termination threshold $\gamma_{th}$.
In each iteration, the algorithm decreases $\gamma$ by a ratio $r$ and calls
the $\textsc{GreedySetCover}$ procedure to solve (\ref{eq:lp-PL-selection}).
The algorithm terminates when $\gamma<\gamma_{th}$. In the initial iterations,
the weight $\gamma$ for the secondary cost is large so that the
algorithm explores solutions which select PLs with less variation.  Later, the
weight $\gamma$ decreases to ensure that the primary objective plays the main
role. Parameters $\gamma_{th}$ and $r$ determine the algorithm's degree of
exploration, which helps avoid local minimum.  In practice, you can choose
small $\gamma_{th}$ and large $r$ if you have enough computation power.

%DONE explain the hard thresholds in HeuristicRefinePl
$\textsc{GreedySetCover}$ uses the ratio of the number of uncovered
windows a PL can cover and the cost $1+\gamma c_{v}$ as heuristics, where
$c_{v}$ is the corresponding \emph{coefficient of variation}.
$\textsc{GreedySetCover}$ will add the PL with the maximum heuristic value to
$\scrP^{F}$ until all windows are covered by the PLs in $\scrP^{F}$. Suppose
the return value of $\textsc{HeuristicRefinePl}$ is $\bx^{*}$. Then, the
refined family of PLs is $\scrP^{F}=\left\{
\bp_{j}^{F}:x_{j}^{*}>0,j=1,\dots,N\right\} $.

\section{Simulation results\label{sec:Network-Simulation}}

Lacking data with annotated anomalies is a common problem for validation
of network anomaly methods. We developed an open source software package
SADIT~\cite{sadit} to provide flow-level datasets with annotated
anomalies. Based on the \emph{fs}-simulator~\cite{Sommers2011},
SADIT simulates the normal and abnormal flows in networks efficiently.

Our simulated network consists of an internal network and
several Internet nodes.
% (Fig.~\ref{fig:scene}). 
The internal network consists
of 8 normal nodes \emph{CT1}-\emph{CT8} and 1 server \emph{SRV} containing
some sensitive information. There are also three Internet nodes
\emph{INT1-INT3} that access the internal network through a gateway
(\emph{GATEWAY}). For all links, the link capacity is $10$ Mb/s and the delay
is 0.01 s. 
All internal and Internet nodes communicate with the \emph{SRV} and there is
no communication between other nodes. The normal flows from all nodes to
\emph{SRV} have the same characteristics. The size of the normal flows follows
a Gaussian distribution $N(m(t),\sigma^{2})$.  The arrival process of flows is
a Poisson process with arrival rate $\lambda(t)$. Both $m(t)$ and $\lambda(t)$
change with time $t$. 

% \begin{figure}
% \begin{centering}
% \includegraphics[width=7cm]{figure/scene}
% \par\end{centering}
% \vspace{-0.2cm}
% \caption{Simulation settings.\label{fig:scene}}
% \vspace{-0.2cm}
% \end{figure}

\begin{figure}[!tp]
% \begin{minipage}[t]{0.45\textwidth}%
\begin{center}
\includegraphics[width=7cm,height=6cm]{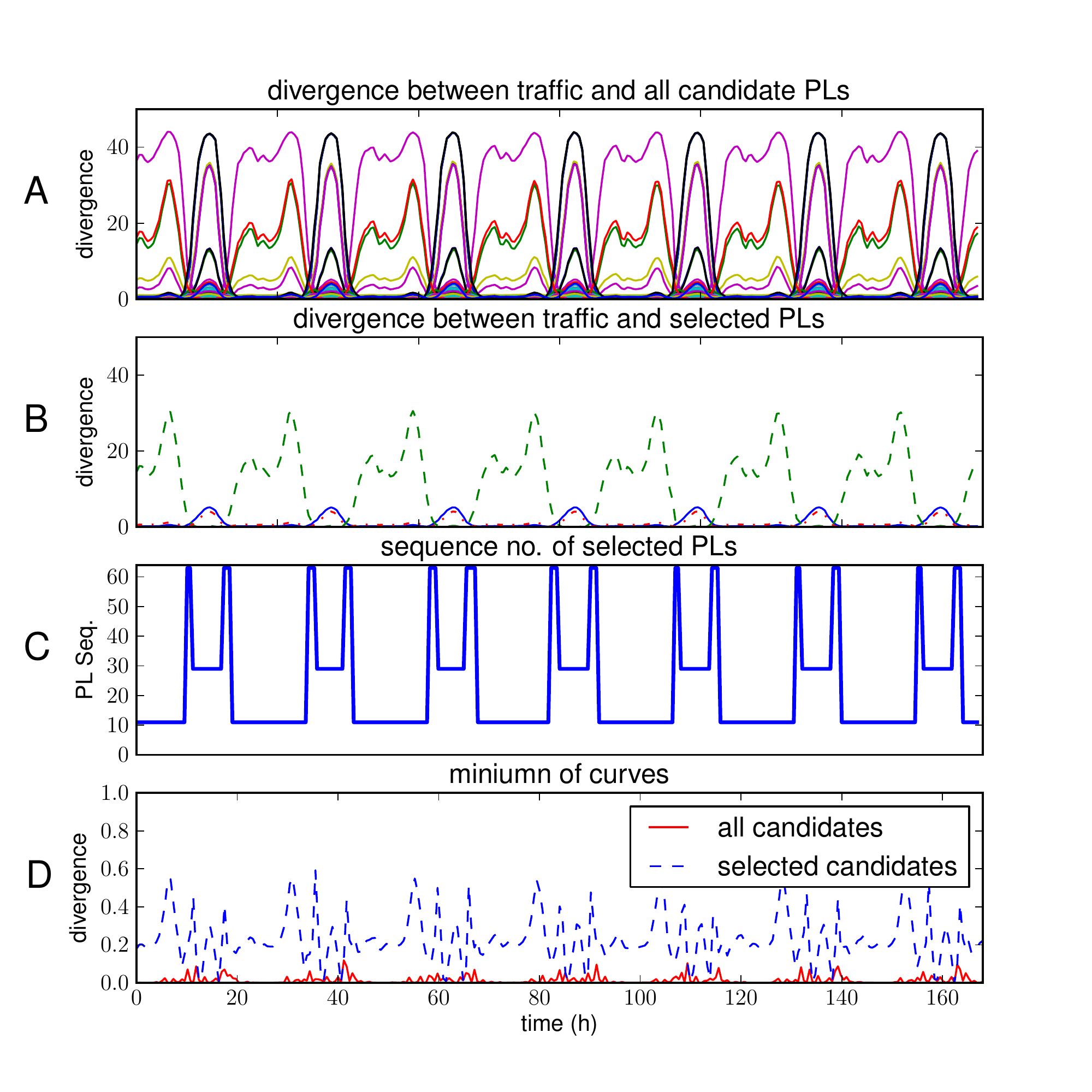}
\par\end{center}

\vspace{-0.5cm}
\caption{Results of PL refinement for \emph{the model-free} method in a network
with \emph{diurnal pattern}\label{fig:pl-ident-mf}. All figures
share the $x$-axis. (A) and (B) plot the divergence of traffic in
each window with all candidate PLs and with selected PLs, respectively.
(C) shows the \emph{active} PL for each window. (D) plots the \emph{generalized
divergence} of traffic in each window with all candidate PLs and selected
PLs.}
\vspace{-0.5cm}
%
% \end{minipage}\hfill{}%

\end{figure}

\begin{figure}[!tp]
\begin{center}
\includegraphics[width=7cm,height=6cm]{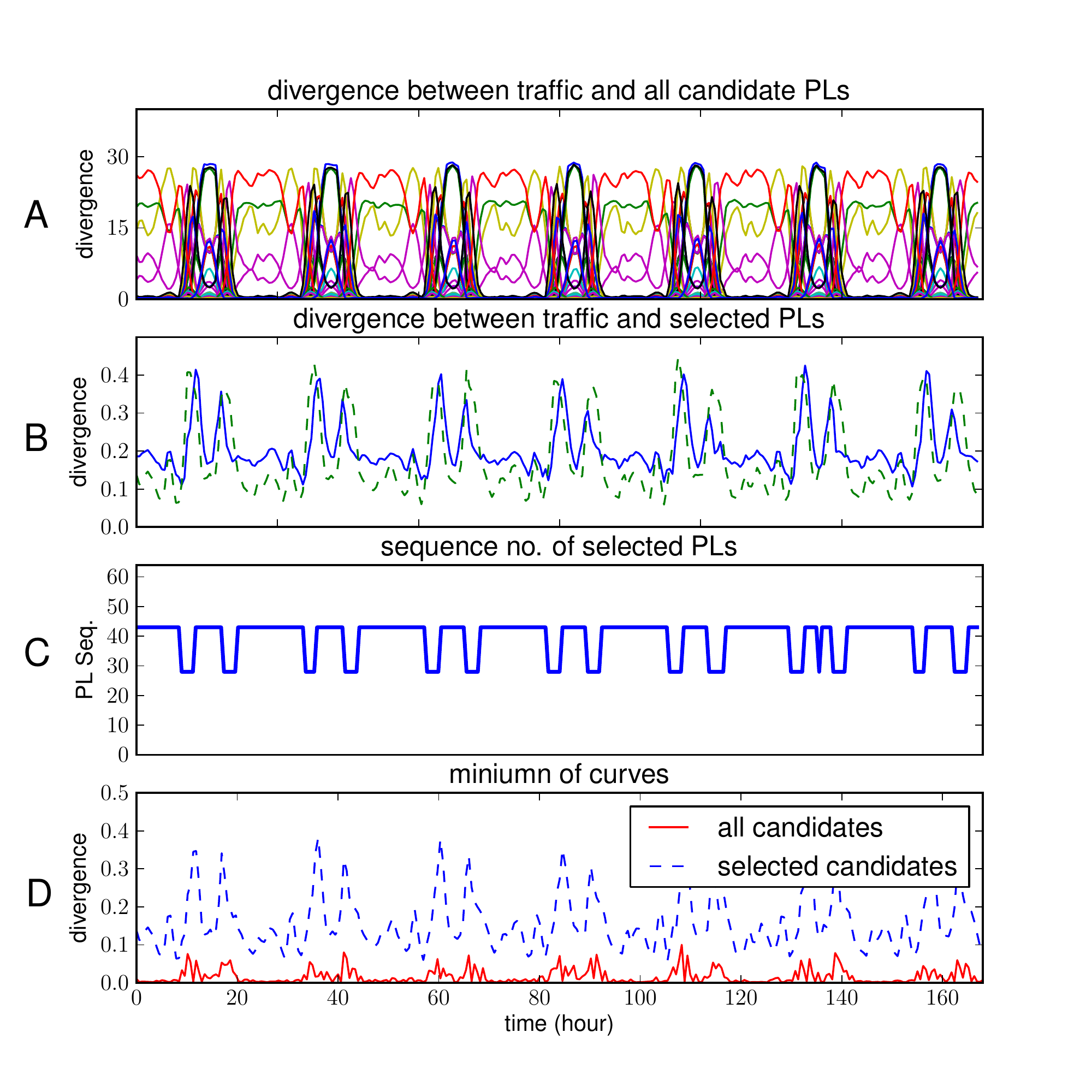}
\par\end{center}

\vspace{-0.5cm}
\caption{Results of PL refinement for \emph{the model-based} method in a network
with \emph{diurnal pattern}\label{fig:pl-ident-mb}. All figures
share the $x$-axis. (A) and (B) plot the divergence of traffic in
each window with all candidate PLs and with selected PLs, respectively.
(C) shows the \emph{active} PL for each window. (D) plots the \emph{generalized
divergence} of traffic in each window with all candidate PLs and selected
PLs.}
\vspace{-0.4cm}
%
% \end{minipage}
\end{figure}

We assume the flow arrival rate and the mean flow size have the same
\emph{diurnal pattern}. Let $p(t)$ be the normalized average traffic to
American social websites~\cite{akamai}, which varies diurnally,
% function shown in
% Fig.~\ref{fig:day-night-traffic}, 
and assume $\lambda(t)=\Lambda p(t)$ and $m(t)=M_pp(t)$, where $\Lambda$
and $M_p$ are the peak arrival rate and the peak mean flow size.  In our
simulation, we set $M_p=4$ Mb, $\sigma^{2}=0.01$, and $\Lambda=0.1$ fps
(flow per second) for all users. Using this \emph{diurnal pattern}, we
generate reference traffic $\calG_{ref}$ for one week~(168 hours) whose
start time is 5 pm. For window aggregation, both the window size $w_{s}$
and the interval $h$ between two consecutive windows is $2,000$ s. The
number of user clusters is $K=2$. The number of quantization levels for
feature 2, 3, 4 are $2$, $2$, and $8$.
An estimation procedure is applied to estimate $t_{d}$ and $t_{p}$. 
The estimate of the period based on flow size is $t_{p}^{3}=24.56$ h with
only $2.3\%$ error. 

\subsection{PL refinement}
For the \emph{model-free} method, there are $64$ candidate
\emph{model-free} PLs. 
% proposed in the estimation stage. 
The \emph{model-free divergence} between each window and each candidate PL is a
periodic function of time, too. Some PLs have smaller divergence during the day
and some others have smaller divergence during the
night~(cf. Fig.~\ref{fig:pl-ident-mf}A).  However, no PL has small
divergence for all windows. $3$ PLs out of the 64 candidates are
selected when the detection threshold is $\lambda=0.6$~(cf.
Fig.~\ref{fig:pl-ident-mf}B). The 3 selected PLs are active during
day, night, and the \emph{transitional} \emph{time}, respectively~(cf.
Fig.~\ref{fig:pl-ident-mf}C for the active PLs of all windows).  For
all windows, the \emph{model-free generalized divergence} between
$\calG_{ref}$ and all candidate PLs is very close to the divergence
between $\calG_{ref}$ and only the selected
PLs~(Fig.~\ref{fig:pl-ident-mf}D). The difference is relatively larger
during the \emph{transitional time} between day and night. This is
because the network is more dynamic during this \emph{transitional
  time}, thus, more PLs are required to represent the network
accurately.

For the \emph{model-based} method, there are 64 candidate
\emph{model-based} PLs, too. Similar to the \emph{model-free }method,
the \emph{model-based} divergence between all candidate PLs and flows in
each window in $\calG_{ref}$ is
periodic~(Fig.~\ref{fig:pl-ident-mb}A) and there is no PL that can
represent all the reference data $\calG_{ref}$. 2 PLs are selected when
$\lambda=0.4$~(Fig.~\ref{fig:pl-ident-mb}B). One PL is active during
the \emph{transitional time} and the other is active during the
\emph{stationary time}, which consists of both day and
night~(Fig.~\ref{fig:pl-ident-mb}C). As before, the divergence between
each $\calG_{ref}^i$ and all candidate PLs is similar to the divergence between
$\calG_{ref}^i$ and just the selected PLs~(Fig.~\ref{fig:pl-ident-mb}D).

The results show that the PL refinement procedure is effective and the
refined family of PLs is meaningful. Each PL in the refined family of
the \emph{ model-free} method corresponds to a ``pattern of normal
behavior,'' whereas, each PL in the refined family of the
\emph{model-based} method describes the transition among the
``patterns''. This information is useful not only for anomaly detection
but also for understanding the normal traffic in dynamic networks.

\subsection{Comparison with vanilla stochastic methods}

\begin{figure}[t]
\begin{centering}
\includegraphics[width=7cm,height=6cm]{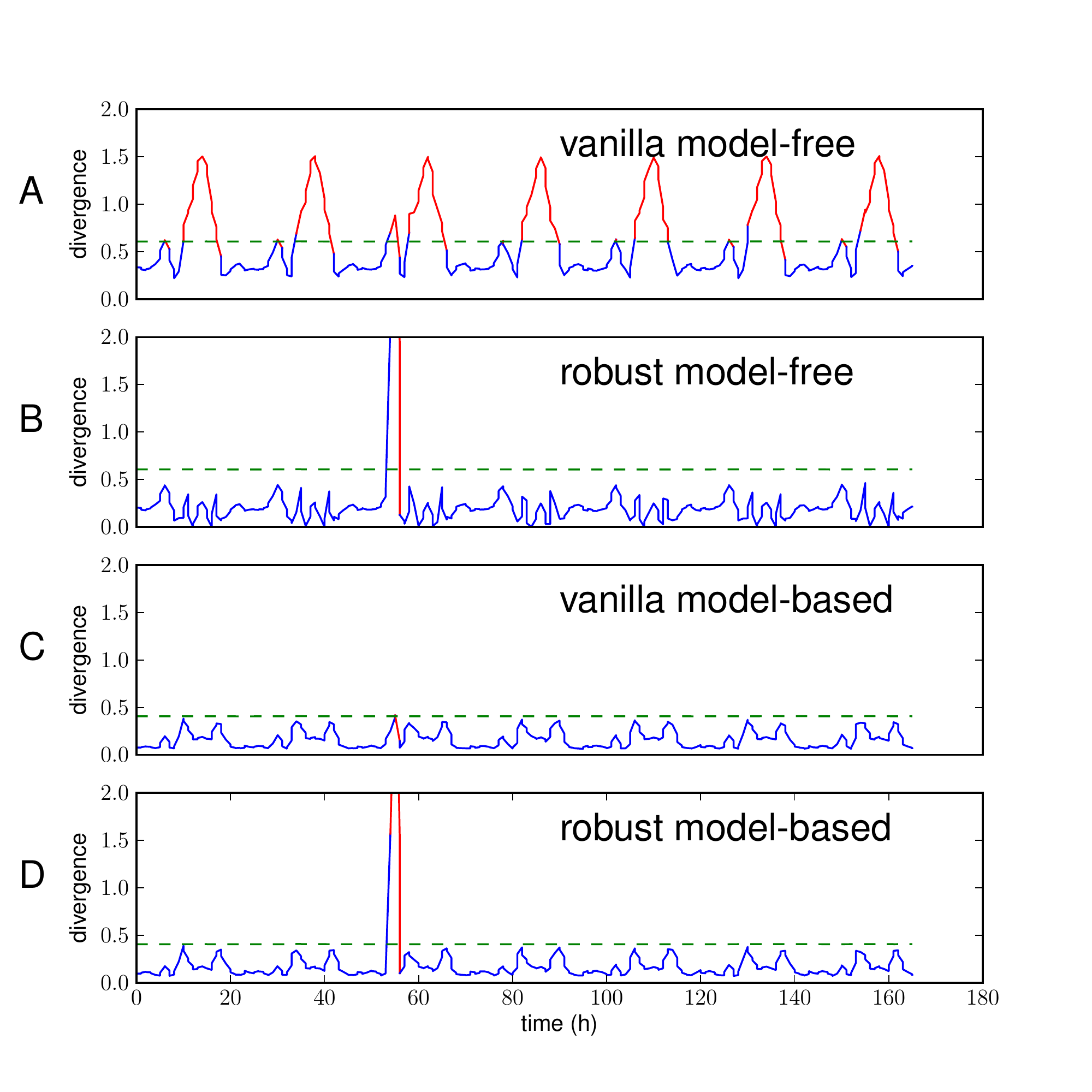}
\par\end{centering}
\vspace{-0.5cm}
\caption{Comparison of vanilla and robust methods.  (A), (B) show detection
    results of vanilla and robust \emph{model-free} methods and (C), (D)
    show detection results of vanilla and robust \emph{model-based}
    methods.  The horizontal lines indicate the detection threshold.
    \label{fig:Comparison-of-Normal-and-Robust}}
\end{figure}

We compared the performance of our robust \emph{model-free} and
\emph{model-based} method with their vanilla
counterparts~(\cite{pas-sma-ton-09,LockeWangPasTech}) in detecting
anomalies.  In the vanilla methods, all reference traffic $\calG_{ref}$ is
used to estimate a single PL.  We used all methods to monitor the server
\emph{SRV} for one week (168 hours). 

We considered an anomaly in which node $CT2$ increases the mean flow
size by $30\%$ at 59h and the increase lasts for 80 minutes before the mean
returns to its normal value. This type of anomaly could be associated with a
situation when attackers try to exfiltrate sensitive information (e.g.,
user accounts and passwords) through SQL injection~\cite{Stampar2013}.

For all methods, the window size is $w_{s}=2000 s$  and the interval
$h=2000 s$. The quantization parameters are equal to those in the
procedure for analyzing the reference traffic $\calG_{ref}$.  The
simulation results show that the robust \emph{model-free} and
\emph{model-based} methods perform better than their vanilla
counterparts for both types of normal traffic
patterns~(Fig.~\ref{fig:Comparison-of-Normal-and-Robust}).

The \emph{diurnal pattern} has large influence on the results of the
vanilla methods. For both the vanilla and the robust \emph{model-free}
methods, the detection threshold $\lambda$ equals 0.6. The vanilla
\emph{model-free} method reports
all night traffic (between 3 am to 11 am) as
anomalies~(Fig.~\ref{fig:Comparison-of-Normal-and-Robust}A).
The reason is that the night traffic is lighter than the day traffic,
so the PL calculated using all of $\calG_{ref}$ is dominated by the
\emph{day pattern}, whereas the \emph{night pattern} is underrepresented.
In contrast, because both the \emph{day} and the \emph{night pattern}
is represented in the refined family of PLs~(Fig.~\ref{fig:pl-ident-mf}B),
the robust \emph{model-free} method is not influenced by the fluctuation
of normal traffic and successfully detects the
anomaly~(Fig.~\ref{fig:Comparison-of-Normal-and-Robust}B). 

The \emph{diurnal pattern} has similar effects on the \emph{model-based}
methods. When the detection threshold $\lambda$ equals 0.4, the anomaly
is barely detectable using the vanilla \emph{model-based}
method~(Fig.~\ref{fig:Comparison-of-Normal-and-Robust}C).
Similar to the vanilla \emph{model-free}
method, the divergence is higher during the \emph{transitional time} because
the \emph{transition pattern} is underrepresented in the PL calculated
using all of $\calG_{ref}$. Again, the robust \emph{model-based }method
is superior because both the \emph{transition pattern} and the \emph{stationary
pattern} are well represented in the refined family of
PLs~(Fig.~\ref{fig:Comparison-of-Normal-and-Robust}D).

\section{Conclusions\label{sec:Conclusions}}

The statistical properties of normal traffic are time-varying for many
networks. We propose a robust \emph{model-free} and a robust
\emph{model-based} method to perform host-based anomaly detection in those
networks.  Our methods can generate a more complete representation of the
normal traffic and are robust to the non-stationarity in networks.

\bibliographystyle{IEEEtran}

% \bibliography{/home/yannisp/Private/bib/abbrev,/home/yannisp/Private/bib/IEEEabrv,/home/yannisp/Private/bib/communications,/home/yannisp/Private/bib/my,/home/yannisp/Private/bib/optimization,/home/yannisp/Private/bib/stochastics,/home/yannisp/Private/bib/various,ano,cyber,web}

% \bibliography{C:/Users/yannisp/Documents/Private/bib/IEEEabrv,C:/Users/yannisp/Documents/Private/bib/abbrev,C:/Users/yannisp/Documents/Private/bib/communications,C:/Users/yannisp/Documents/Private/bib/my,C:/Users/yannisp/Documents/Private/bib/optimization,ano,cyber,web}

% \bibliography{ano,cyber,web}
\bibliography{anomaly_med-2014.bbl}

\end{document}

%% file: mymacros_2cln.tex
\newcommand{\be}[1]{\begin{equation}\label{#1}}
\newcommand{\benon}{\begin{equation*}}  
\newcommand{\bemuln}[1]{\begin{multline}\label{#1}}
\newcommand{\bemul}{\begin{multline*}}
\newcommand{\bee}{\begin{eqnarray*}}
\newcommand{\eee}{\end{eqnarray*}}
\newcommand{\been}[1]{\begin{eqnarray}\label{#1}}
\newcommand{\eeen}{\end{eqnarray}}
\newcommand{\began}[1]{\begin{gather}\label{#1}}
\newcommand{\bega}{\begin{gather*}}
\newcommand{\bealn}[1]{\begin{align}\label{#1}}
\newcommand{\beal}{\begin{align*}}
\newcommand{\bealatn}[2]{\begin{alignat}{#1}\label{#2}}
\newcommand{\bealat}{\begin{alignat*}}
\newcommand{\bexalatn}[1]{\begin{xalignat}\label{#1}}
\newcommand{\bexalat}{\begin{xalignat*}}
\newcommand{\ra}{\rightarrow}
\newcommand{\Ra}{\Rightarrow}

\newcommand{\diag}{\text{diag}}
\newcommand{\tsum}{\textstyle\sum}
\newcommand{\dsum}{\displaystyle\sum}
\def\argmax{\mathop{\rm arg\,max}}

\newcommand{\mb}{\mathbf}
\newcommand{\bsy}{\boldsymbol}
\newcommand{\mn}{\mathnormal}
\newcommand{\mbb}{\mathbb}
\newcommand{\bosy}{\boldsymbol}

{\theoremstyle{plain} \newtheorem{thm}{Theorem}[section]

}
{\theoremstyle{break} \theorembodyfont{\it}
\newtheorem{defi}{Definition}
 }

%% file: lpsymbols.tex
\def\bb{{\mathbf b}}
\def\bc{{\mathbf c}}
\def\bd{{\mathbf d}}
\def\bg{{\mathbf g}}
\def\bh{{\mathbf h}}
\def\bi{{\mathbf i}}
\def\bl{{\mathbf l}}
\def\bm{{\mathbf m}}
\def\bp{{\mathbf p}}
\def\bq{{\mathbf q}}
\def\br{{\mathbf r}}
\def\bs{{\mathbf s}}
\def\bt{{\mathbf t}}
\def\bu{{\mathbf u}}
\def\bv{{\mathbf v}}
\def\bw{{\mathbf w}}
\def\bx{{\mathbf x}}
\def\by{{\mathbf y}}
\def\bz{{\mathbf z}}
\def\bA{{\mathbf A}}
\def\bB{{\mathbf B}}
\def\bC{{\mathbf C}}
\def\bD{{\mathbf D}}
\def\bE{{\mathbf E}}
\def\bF{{\mathbf F}}
\def\bG{{\mathbf G}}
\def\bH{{\mathbf H}}
\def\bI{{\mathbf I}}
\def\bK{{\mathbf K}}
\def\bL{{\mathbf L}}
\def\bM{{\mathbf M}}
\def\bN{{\mathbf N}}
\def\bO{{\mathbf O}}
\def\bP{{\mathbf P}}
\def\bQ{{\mathbf Q}}
\def\bR{{\mathbf R}}

\def\bU{{\mathbf U}}
\def\bT{{\mathbf T}}

\def\bV{{\mathbf V}}
\def\bW{{\mathbf W}}
\def\bZ{{\mathbf Z}}

\def\texitem#1{\par\smallskip\noindent\hangindent 25pt
               \hbox to 25pt {\hss #1 ~}\ignorespaces}

\newcommand{\bzero}{{\mathbf{0}}}

\newcommand{\scrA}{\mathcal{A}}
\newcommand{\scrB}{\mathcal{B}}
\newcommand{\scrC}{\mathcal{C}}
\newcommand{\scrD}{\mathcal{D}}
\newcommand{\scrE}{\mathcal{E}}
\newcommand{\scrF}{\mathcal{F}}
\newcommand{\scrG}{\mathcal{G}}
\newcommand{\scrh}{\mathcal{h}}
\newcommand{\scrH}{\mathcal{H}}
\newcommand{\scrI}{\mathcal{I}}
\newcommand{\scrJ}{\mathcal{J}}
\newcommand{\scrK}{\mathcal{K}}
\newcommand{\scrL}{\mathcal{L}}
\newcommand{\scrM}{\mathcal{M}}
\newcommand{\scrN}{\mathcal{N}}
\newcommand{\scrO}{\mathcal{O}}
\newcommand{\scrP}{\mathcal{P}}
\newcommand{\scrQ}{\mathcal{Q}}
\newcommand{\scrR}{\mathcal{R}}
\newcommand{\scrS}{\mathcal{S}}
\newcommand{\scrT}{\mathcal{T}}
\newcommand{\scrU}{\mathcal{U}}
\newcommand{\scrV}{\mathcal{V}}
\newcommand{\scrW}{\mathcal{W}}
\newcommand{\scrX}{\mathcal{X}}
\newcommand{\scrY}{\mathcal{Y}}

\newcommand{\balpha}{\boldsymbol{\alpha}}

\newcommand{\bgamma}{\boldsymbol{\gamma}}
\newcommand{\bdelta}{{\boldsymbol{\delta}}}
\newcommand{\bepsilon}{\boldsymbol{\epsilon}}
\newcommand{\bzeta}{\boldsymbol{\zeta}}
\newcommand{\btheta}{\boldsymbol{\theta}}
\newcommand{\bkappa}{\boldsymbol{\kappa}}
\newcommand{\blambda}{\boldsymbol{\lambda}}
\newcommand{\bmu}{\boldsymbol{\mu}}
\newcommand{\bnu}{{\boldsymbol{\nu}}}
\newcommand{\bpi}{{\boldsymbol{\pi}}}
\newcommand{\brho}{{\boldsymbol{\rho}}}
\newcommand{\bsigma}{{\boldsymbol{\sigma}}}
\newcommand{\bSigma}{{\boldsymbol{\Sigma}}}
\newcommand{\bphi}{{\boldsymbol{\phi}}}
\newcommand{\bPhi}{\boldsymbol{\Phi}}
\newcommand{\bxi}{\boldsymbol{\xi}}
\newcommand{\bpsi}{\boldsymbol{\psi}}
\newcommand{\bomega}{\boldsymbol{\omega}}
\newcommand{\bOmega}{\boldsymbol{\Omega}}

%% file: wj_macros.tex
\global\long\def\btheta{\boldsymbol{\theta}}
 \global\long\def\bth{\boldsymbol{\theta}}
 \global\long\def\bpsi{\boldsymbol{\psi}}
 % \global\long\def\mbb#1{#1}
\global\long\def\ra{\rightarrow}
 \global\long\def\Ra{\Rightarrow}
 \global\long\def\diag{\text{diag}}
 \global\long\def\tsum{{\textstyle \sum}}
 \global\long\def\dsum{{\displaystyle \sum}}
 \global\long\def\tprod{{\textstyle \prod}}
 \global\long\def\argmax{\mathop{\rm arg\, max}}
 \global\long\def\mb{\mathbf{}}
 \global\long\def\bsy{\boldsymbol{}}
 \global\long\def\mn{\mathnormal{}}
 \global\long\def\bosy{\boldsymbol{}}
 \global\long\def\bj{\mathbf{j}}
 \global\long\def\bb{\mathbf{b}}
 \global\long\def\bc{\mathbf{c}}
 \global\long\def\bd{\mathbf{d}}
 \global\long\def\bbf{\mathbf{f}}
 \global\long\def\f{\mathbf{f}}
 \global\long\def\bg{\mathbf{g}}
 \global\long\def\bg{\mathbf{g}}
 \global\long\def\bh{\mathbf{h}}
 \global\long\def\bi{\mathbf{i}}
 \global\long\def\bl{\mathbf{l}}
 \global\long\def\bm{\mathbf{m}}
 \global\long\def\bp{\mathbf{p}}
 \global\long\def\bq{\mathbf{q}}
 \global\long\def\br{\mathbf{r}}
 \global\long\def\bs{\mathbf{s}}
 \global\long\def\bt{\mathbf{t}}
 \global\long\def\bu{\mathbf{u}}
 \global\long\def\bv{\mathbf{v}}
 \global\long\def\bw{\mathbf{w}}
 \global\long\def\bx{\mathbf{x}}
 \global\long\def\by{\mathbf{y}}
 \global\long\def\bz{\mathbf{z}}
 \global\long\def\bA{\mathbf{A}}
 \global\long\def\bB{\mathbf{B}}
 \global\long\def\bC{\mathbf{C}}
 \global\long\def\bD{\mathbf{D}}
 \global\long\def\bE{\mathbf{E}}
 \global\long\def\bF{\mathbf{F}}
 \global\long\def\bG{\mathbf{G}}
 \global\long\def\bH{\mathbf{H}}
 \global\long\def\bI{\mathbf{I}}
 \global\long\def\bJ{\mathbf{J}}
 \global\long\def\bL{\mathbf{L}}
 \global\long\def\bM{\mathbf{M}}
 \global\long\def\bN{\mathbf{N}}
 \global\long\def\bO{\mathbf{O}}
 \global\long\def\bP{\mathbf{P}}
 \global\long\def\bQ{\mathbf{Q}}
 \global\long\def\bR{\mathbf{R}}
 \global\long\def\bT{\mathbf{T}}
 \global\long\def\bU{\mathbf{U}}
 \global\long\def\bV{\mathbf{V}}
 \global\long\def\bW{\mathbf{W}}
 \global\long\def\bZ{\mathbf{X}}
 \global\long\def\bzero{\mathbf{0}}
 \global\long\def\1{\mathbf{1}}
 % \global\long\def\calA{\mathcal{A}}
% \global\long\def\calH{\mathcal{H}}
% \global\long\def\calL{\mathcal{L}}

\global\long\def\calA{\mathcal{A}}
 \global\long\def\calB{\mathcal{B}}
 \global\long\def\calC{\mathcal{C}}
 \global\long\def\calD{\mathcal{D}}
 \global\long\def\calE{\mathcal{E}}
 \global\long\def\calF{\mathcal{F}}
 \global\long\def\calG{\mathcal{G}}
 \global\long\def\calH{\mathcal{H}}
 \global\long\def\calI{\mathcal{I}}
 \global\long\def\calJ{\mathcal{J}}
 \global\long\def\calK{\mathcal{K}}
 \global\long\def\calL{\mathcal{L}}
 \global\long\def\calM{\mathcal{M}}
 \global\long\def\calN{\mathcal{N}}
 \global\long\def\calO{\mathcal{O}}
 \global\long\def\calP{\mathcal{P}}
 \global\long\def\calR{\mathcal{R}}
 \global\long\def\calS{\mathcal{S}}
 \global\long\def\calQ{\mathcal{Q}}
 \global\long\def\calU{\mathcal{U}}
 \global\long\def\calV{\mathcal{V}}
 \global\long\def\calW{\mathcal{W}}
 \global\long\def\calX{\mathcal{X}}
 \global\long\def\calY{\mathcal{Y}}
 \global\long\def\calZ{\mathcal{Z}}

\global\long\def\bcalA{\boldsymbol{\mathcal{A}}}
 \global\long\def\bcalB{\boldsymbol{\mathcal{B}}}
 \global\long\def\bcalC{\boldsymbol{\mathcal{C}}}
 \global\long\def\bcalD{\boldsymbol{\mathcal{D}}}
 \global\long\def\bcalE{\boldsymbol{\mathcal{E}}}
 \global\long\def\bcalF{\boldsymbol{\mathcal{F}}}
 \global\long\def\bcalG{\boldsymbol{\mathcal{G}}}
 \global\long\def\bcalH{\boldsymbol{\mathcal{H}}}
 \global\long\def\bcalI{\boldsymbol{\mathcal{I}}}
 \global\long\def\bcalJ{\boldsymbol{\mathcal{J}}}
 \global\long\def\bcalK{\boldsymbol{\mathcal{K}}}
 \global\long\def\bcalL{\boldsymbol{\mathcal{L}}}
 \global\long\def\bcalM{\boldsymbol{\mathcal{M}}}
 \global\long\def\bcalN{\boldsymbol{\mathcal{N}}}
 \global\long\def\bcalO{\boldsymbol{\mathcal{O}}}
 \global\long\def\bcalP{\boldsymbol{\mathcal{P}}}
 \global\long\def\bcalR{\boldsymbol{\mathcal{R}}}
 \global\long\def\bcalS{\boldsymbol{\mathcal{S}}}
 \global\long\def\bcalQ{\boldsymbol{\mathcal{Q}}}
 \global\long\def\bcalU{\boldsymbol{\mathcal{U}}}
 \global\long\def\bcalV{\boldsymbol{\mathcal{V}}}
 \global\long\def\bcalW{\boldsymbol{\mathcal{W}}}
 \global\long\def\bcalX{\boldsymbol{\mathcal{X}}}
 \global\long\def\bcalY{\boldsymbol{\mathcal{Y}}}
 \global\long\def\bcalZ{\boldsymbol{\mathcal{Z}}}

% \global\long\def\scrA{\mathcal{A}}
% \global\long\def\scrB{\mathcal{B}}
% \global\long\def\scrC{\mathcal{C}}
% \global\long\def\scrD{\mathcal{D}}
% \global\long\def\scrE{\mathcal{E}}
% \global\long\def\scrF{\mathcal{F}}
% \global\long\def\scrG{\mathcal{G}}
% \global\long\def\scrh{\mathcal{h}}
% \global\long\def\scrH{\mathcal{H}}
% \global\long\def\scrI{\mathcal{I}}
% \global\long\def\scrJ{\mathcal{J}}
% \global\long\def\scrK{\mathcal{K}}
% \global\long\def\scrL{\mathcal{L}}
% \global\long\def\scrM{\mathcal{M}}
% \global\long\def\scrN{\mathcal{N}}
% \global\long\def\scrO{\mathcal{O}}
% \global\long\def\scrP{\mathcal{P}}
% \global\long\def\scrQ{\mathcal{Q}}
% \global\long\def\scrR{\mathcal{R}}
% \global\long\def\scrS{\mathcal{S}}
% \global\long\def\scrT{\mathcal{T}}
% \global\long\def\scrU{\mathcal{U}}
% \global\long\def\scrV{\mathcal{V}}
% \global\long\def\scrW{\mathcal{W}}
% \global\long\def\scrX{\mathcal{X}}
% \global\long\def\scrY{\mathcal{Y}}
% \global\long\def\scrT{\mathcal{T}}

\global\long\def\scrA{\mathscr{A}}
 \global\long\def\scrB{\mathscr{B}}
 \global\long\def\scrC{\mathscr{C}}
 \global\long\def\scrD{\mathscr{D}}
 \global\long\def\scrE{\mathscr{E}}
 \global\long\def\scrF{\mathscr{F}}
 \global\long\def\scrG{\mathscr{G}}
 \global\long\def\scrh{\mathscr{h}}
 \global\long\def\scrH{\mathscr{H}}
 \global\long\def\scrI{\mathscr{I}}
 \global\long\def\scrJ{\mathscr{J}}
 \global\long\def\scrK{\mathscr{K}}
 \global\long\def\scrL{\mathscr{L}}
 \global\long\def\scrM{\mathscr{M}}
 \global\long\def\scrN{\mathscr{N}}
 \global\long\def\scrO{\mathscr{O}}
 \global\long\def\scrP{\mathscr{P}}
 \global\long\def\scrQ{\mathscr{Q}}
 \global\long\def\scrR{\mathscr{R}}
 \global\long\def\scrS{\mathscr{S}}
 \global\long\def\scrT{\mathscr{T}}
 \global\long\def\scrU{\mathscr{U}}
 \global\long\def\scrV{\mathscr{V}}
 \global\long\def\scrW{\mathscr{W}}
 \global\long\def\scrX{\mathscr{X}}
 \global\long\def\scrY{\mathscr{Y}}
 \global\long\def\scrZ{\mathscr{Z}}

\global\long\def\balpha{\boldsymbol{\alpha}}
 \global\long\def\bgamma{\boldsymbol{\gamma}}
 \global\long\def\bdelta{{\boldsymbol{\delta}}}
 \global\long\def\bepsilon{\boldsymbol{\epsilon}}
 \global\long\def\bzeta{\boldsymbol{\zeta}}
 \global\long\def\btheta{\boldsymbol{\theta}}
 \global\long\def\bth{\boldsymbol{\theta}}
 \global\long\def\bkappa{\boldsymbol{\kappa}}
 \global\long\def\blambda{\boldsymbol{\lambda}}
 \global\long\def\bmu{\boldsymbol{\mu}}
 \global\long\def\bnu{\boldsymbol{\nu}}
 \global\long\def\bpi{\boldsymbol{\pi}}
 \global\long\def\brho{\boldsymbol{\rho}}
 \global\long\def\bsigma{\boldsymbol{\sigma}}
 \global\long\def\bSigma{\boldsymbol{\Sigma}}

\global\long\def\bxi{\boldsymbol{\xi}}
 \global\long\def\bXi{\boldsymbol{\Xi}}
 \global\long\def\bpsi{\boldsymbol{\psi}}
 \global\long\def\bomega{\boldsymbol{\omega}}
 \global\long\def\bOmega{\boldsymbol{\Omega}}

\global\long\def\bTheta{\boldsymbol{\Theta}}

\global\long\def\bphi{\boldsymbol{\phi}}

\global\long\def\bPhi{\boldsymbol{\Phi}}

\global\long\def\bzero{\mathbf{0}}

\global\long\def\mb#1{\mathbb{#1}}

\global\long\def\mc#1{\mathcal{#1}}

\global\long\def\bK{\mathbf{K}}

\global\long\def\bPsi{\boldsymbol{\Psi}}

\global\long\def\mbb{\mathbb{}}
 \global\long\def\lt{\ensuremath{}}
 \global\long\def\rt{\ensuremath{}}

%% file: HeuristicRefine.tex
\newcommand{\algorithmicbreak}{\textbf{break}}
\newcommand{\Continue}{\textbf{continue}}
\newcommand{\Break}{\State \algorithmicbreak}
\begin{algorithmic}
\Function {HeuristicRefinePl}{$\bA$, $\bc_v$, r, $\gamma_{th}$}
    % \State\textbf{Init}: $k\gets 0$
    % \State Calculate $\bA = \{a_{ij}: a_{ij} = 1(D_{ij} < \lambda)\}$
    % \For {$i = 1 \to M$, $j = 1 \to N$}
        % \For {$j = 1 \to N$}
        % \If {$\bD$[i][j] < $\lambda$}
        %     $\bA$[i][j] = 1
        % \Else
        %     \quad $\bA$[i][j] = 0
        % \EndIf
        % \EndFor
    % \EndFor
    \State \textbf{Init}: bestCost := $\infty$, $\gamma := 1$, $\bx^*:=0$
    \While {$\gamma \geq \gamma_{th}$ }
        % \State {$k \gets k+1$}
        \State 
        % \State {$\gamma_{k} \gets r \gamma_{k-1}$}
        $\bx$ := \Call {GreedySolve}{$\bA$, $\gamma$, $\bc_v$},
        $\gamma := r \gamma$
        \If {$\mathbf{1}^{'} \bx + \gamma_{th} \bc_v^{'} \bx <$ bestCost}
        \State bestCost :=  $\mathbf{1}^{'} \bx + \gamma_{th} \bc_v^{'} \bx$
            \State $\bx^* := \bx$
        \EndIf
    \EndWhile
    % \State \Return $\bx* = \arg\max_{\bx^k} \sum_{j}\bx^k[j]$
    \State \Return $\bx^*$
\EndFunction

\Function {GreedySetCover}{$\bA$, $\gamma$, $\bc_v$}
    \State\textbf{Init}: $\bx^0 := \bzero$, $C := \varnothing$
    \While {$|C| < M$}
    % \For {$j \mbox{ such that }\bx[j] = 0$}
    % \State $\bs[j] \gets \frac{\sum_{i\notin C} a_{ij}}{1 + \gamma \bc_v[j]} $
    % \EndFor
    % \State $j^+ \gets \arg\max_{j:\bx[j]=0} \bs[j]$
    \State $j^+ := \arg\max_{j:\bx[j]=0} \frac{\sum_{i\notin C} a_{ij}}{1 + \gamma \bc_v[j]} $
    \State $\bx[j^+] := 1$, $C := C \cup \{i : a_{ij^+} = 1\}$
    \EndWhile
    \State \Return $\bx$
    % \State \Return $\bx$, $\mathbf{1}^T \bx + \gamma \bc_v^T \bx$
\EndFunction

\end{algorithmic}

%% file: anomaly_med-2014.bbl
\begin{thebibliography}{10}
\providecommand{\url}[1]{#1}
\csname url@samestyle\endcsname
\providecommand{\newblock}{\relax}
\providecommand{\bibinfo}[2]{#2}
\providecommand{\BIBentrySTDinterwordspacing}{\spaceskip=0pt\relax}
\providecommand{\BIBentryALTinterwordstretchfactor}{4}
\providecommand{\BIBentryALTinterwordspacing}{\spaceskip=\fontdimen2\font plus
\BIBentryALTinterwordstretchfactor\fontdimen3\font minus
  \fontdimen4\font\relax}
\providecommand{\BIBforeignlanguage}[2]{{%
\expandafter\ifx\csname l@#1\endcsname\relax
\typeout{** WARNING: IEEEtran.bst: No hyphenation pattern has been}%
\typeout{** loaded for the language `#1'. Using the pattern for}%
\typeout{** the default language instead.}%
\else
\language=\csname l@#1\endcsname
\fi
#2}}
\providecommand{\BIBdecl}{\relax}
\BIBdecl

\bibitem{roesch1999snort}
M.~Roesch \emph{et~al.}, ``Snort-lightweight intrusion detection for
  networks,'' in \emph{Proceedings of the 13th USENIX conference on System
  administration}.\hskip 1em plus 0.5em minus 0.4em\relax Seattle, Washington,
  1999, pp. 229--238.

\bibitem{paxson1999bro}
V.~Paxson, ``Bro: a system for detecting network intruders in real-time,''
  \emph{Computer networks}, vol.~31, no.~23, pp. 2435--2463, 1999.

\bibitem{barford2002signal}
P.~Barford, J.~Kline, D.~Plonka, and A.~Ron, ``A signal analysis of network
  traffic anomalies,'' in \emph{Proceedings of the 2nd ACM SIGCOMM Workshop on
  Internet measurment}.\hskip 1em plus 0.5em minus 0.4em\relax ACM, 2002, pp.
  71--82.

\bibitem{Lu2009}
W.~Lu and A.~a. Ghorbani, ``{Network Anomaly Detection Based on Wavelet
  Analysis},'' \emph{EURASIP Journal on Advances in Signal Processing}, vol.
  2009, no.~1, p. 837601, 2009.

\bibitem{pas-sma-ton-09}
I.~C. Paschalidis and G.~Smaragdakis, ``Spatio-temporal network anomaly
  detection by assessing deviations of empirical measures,'' \emph{Networking,
  IEEE/ACM Transactions on}, vol.~17, no.~3, pp. 685--697, 2009.

\bibitem{lippmann2000evaluating}
R.~P. Lippmann, D.~J. Fried, I.~Graf, J.~W. Haines, K.~R. Kendall, D.~McClung,
  D.~Weber, S.~E. Webster, D.~Wyschogrod, R.~K. Cunningham \emph{et~al.},
  ``Evaluating intrusion detection systems: The 1998 darpa off-line intrusion
  detection evaluation,'' in \emph{DARPA Information Survivability Conference
  and Exposition, 2000. DISCEX'00. Proceedings}, vol.~2.\hskip 1em plus 0.5em
  minus 0.4em\relax IEEE, 2000, pp. 12--26.

\bibitem{deze2}
A.~Dembo and O.~Zeitouni, \emph{{Large Deviations Techniques and
  Applications}}, 2nd~ed.\hskip 1em plus 0.5em minus 0.4em\relax
  NY:Spring-Verlag, 1998.

\bibitem{Neal-2010}
N.~Leavitt, ``{Network-usage changes push internet traffic to the edge},''
  \emph{Computer}, pp. 13--15, 2010.

\bibitem{thompson1997wide}
K.~Thompson, G.~J. Miller, and R.~Wilder, ``Wide-area {Internet} traffic
  patterns and characteristics,'' \emph{Network, IEEE}, vol.~11, no.~6, pp.
  10--23, 1997.

\bibitem{King2013}
A.~King, B.~Huffaker, A.~Dainotti, and K.~C. Claffy, ``{A coordinated view of
  the temporal evolution of large-scale Internet events},'' \emph{Computing},
  pp. 53--65, Jan. 2013.

\bibitem{Sandvine2013}
Sandvine, ``Global internet phenomena report,''
  \url{https://www.sandvine.com/downloads/general/global-internet-phenomena/2013/sandvine-global-internet-phenomena-report-1h-2013.pdf},
  2013.

\bibitem{hoef65}
W.~Hoeffding, ``{Asymptotically optimal tests for multinomial distributions},''
  \emph{Ann. Math. Statist.}, vol.~36, pp. 369--401, 1965.

\bibitem{mainlocalization}
I.~C. Paschalidis and D.~Guo, ``{Robust and distributed stochastic localization
  in sensor networks: Theory and experimental results},'' \emph{ACM
  Transactions on Sensor Networks}, vol.~5, no.~4, 2009.

\bibitem{netflow}
\mbox{Cisco} System, ``Cisco netflow,''
  \url{http://en.wikipedia.org/wiki/NetFlow}, 2012.

\bibitem{sadit}
J.~Wang, ``{SADIT: Systematic Anomaly Detection of Internet Traffic},''
  \url{http://people.bu.edu/wangjing/open-source/sadit/html/index.html}, 2012.

\bibitem{Sommers2011}
J.~Sommers, R.~Bowden, B.~Eriksson, P.~Barford, M.~Roughan, and N.~Duffield,
  ``{Efficient network-wide flow record generation},'' pp. 2363--2371, 2011.

\bibitem{akamai}
A.~Technologies, ``{The Net Usage Index by Industry},''
  \url{http://www.akamai.com/html/technology/nui/industry/index.html}, 2013.

\bibitem{LockeWangPasTech}
R.~Locke, J.~Wang, and I.~Paschalidis, ``Anomaly detection techniques for data
  exfiltration attempts,'' Center for Information \& Systems Engineering,
  Boston University, 8 Saint Mary's Street, Brookline, MA, Tech. Rep.
  2012-JA-0001, June 2012.

\bibitem{Stampar2013}
\BIBentryALTinterwordspacing
M.~Stampar, ``{Data Retrieval over DNS in SQL Injection Attacks},'' \emph{arXiv
  preprint arXiv:1303.3047}, 2013. [Online]. Available:
  \url{http://arxiv.org/abs/1303.3047}
\BIBentrySTDinterwordspacing

\end{thebibliography}
